\RequirePackage{latexml}
\iflatexml
\documentclass{article}
\usepackage{amsmath,amssymb,amsthm,graphicx,booktabs,array,multirow}
\usepackage[numbers,square,sort&compress]{natbib}
\usepackage{hyperref,xcolor,authblk}
\definecolor{QKPrimary}{HTML}{19364A}

\newenvironment{qkfrontmatter}{}{}
\newcommand{\qkkeywords}[1]{\par\noindent\textbf{Keywords: }#1\par}
\newcommand{\qkbibliography}[1]{
}
\newcommand{\qkappendix}{\appendix}
\newcommand{\FloatBarrier}{}
\newcommand{\Needspace}[1]{}
\newcommand{\balance}{}
\newcommand{\nobalance}{}
\providecommand{\pdfbookmark}[3][]{}
\makeatletter
\let\ps@qkbody\ps@plain
\makeatother

\else
  \documentclass[paper]{qiankunnet}
  \AtBeginEnvironment{thebibliography}{\nobalance}
  \makeatletter
  \patchcmd{\@lbibitem}{\bibitem@fin}{\bibitem@fin\ifstrequal{#2}{sun2020pyscf}{\newpage}{}}{}{\ClassError{arxiv}{Bibliography layout patch failed}{}}
  \makeatother
\fi
\newcommand{\R}{\mathbf R}
\newcommand{\nn}{\mathbf n}
\newcommand{\E}{\mathbb E}
\newcommand{\ket}[1]{\lvert #1\rangle}

\newcommand{\result}[1]{\ifcsname result@#1\endcsname\csname result@#1\endcsname\else\PackageError{manuscript}{Missing validated result: #1}{Populate the validated generated results before compiling.}\fi}
\expandafter\def\csname result@c2h4-occupation-mae\endcsname{0.0009}
\expandafter\def\csname result@co2-quadrupole-mae\endcsname{0.0403}
\expandafter\def\csname result@current-co-r-energy-film-seed11\endcsname{0.277}
\expandafter\def\csname result@current-co-r-energy-film-seed23\endcsname{0.194}
\expandafter\def\csname result@current-co-r-energy-film-seed37\endcsname{0.382}
\expandafter\def\csname result@current-co-r-energy-input-seed11\endcsname{0.364}
\expandafter\def\csname result@current-co-r-energy-input-seed23\endcsname{0.438}
\expandafter\def\csname result@current-co-r-energy-input-seed37\endcsname{0.443}
\expandafter\def\csname result@current-co-r-film-seed11\endcsname{0.328}
\expandafter\def\csname result@current-co-r-film-seed23\endcsname{0.333}
\expandafter\def\csname result@current-co-r-film-seed37\endcsname{0.247}
\expandafter\def\csname result@current-co-r-input-seed11\endcsname{0.292}
\expandafter\def\csname result@current-co-r-input-seed23\endcsname{0.414}
\expandafter\def\csname result@current-co-r-input-seed37\endcsname{0.575}
\expandafter\def\csname result@current-n2-aligned-seed11\endcsname{0.049}
\expandafter\def\csname result@current-n2-aligned-seed23\endcsname{0.085}
\expandafter\def\csname result@current-n2-aligned-seed37\endcsname{0.053}
\expandafter\def\csname result@current-n2-raw-seed11\endcsname{36.806}
\expandafter\def\csname result@current-n2-raw-seed23\endcsname{33.609}
\expandafter\def\csname result@current-n2-raw-seed37\endcsname{35.027}
\expandafter\def\csname result@final-c2h4-query-mae\endcsname{0.946}
\expandafter\def\csname result@final-c2h4-query-max\endcsname{1.233}
\expandafter\def\csname result@final-co-query-mae\endcsname{0.253}
\expandafter\def\csname result@final-co-query-max\endcsname{0.585}
\expandafter\def\csname result@final-co2-query-mae\endcsname{0.535}
\expandafter\def\csname result@final-co2-query-max\endcsname{0.895}
\expandafter\def\csname result@final-h4-query-mae\endcsname{0.076}
\expandafter\def\csname result@final-h4-query-max\endcsname{0.120}
\expandafter\def\csname result@final-n2-query-mae\endcsname{0.049}
\expandafter\def\csname result@final-n2-query-max\endcsname{0.095}
\expandafter\def\csname result@final-nh3-query-mae\endcsname{1.036}
\expandafter\def\csname result@final-nh3-query-max\endcsname{1.456}
\expandafter\def\csname result@nh3-dipole-mae\endcsname{0.0058}
\expandafter\def\csname result@prod-c2h4-mae\endcsname{0.919}
\expandafter\def\csname result@prod-co-mae\endcsname{0.197}
\expandafter\def\csname result@prod-co2-mae\endcsname{0.591}
\expandafter\def\csname result@prod-h4-mae\endcsname{0.088}
\expandafter\def\csname result@prod-n2-mae\endcsname{0.036}
\expandafter\def\csname result@prod-nh3-mae\endcsname{0.872}

\expandafter\def\csname result@additional-h4-axis-boundary-mae_mha\endcsname{0.061}
\expandafter\def\csname result@additional-h4-axis-boundary-maximum_absolute_error_mha\endcsname{0.108}
\expandafter\def\csname result@additional-h4-axis-exterior-mae_mha\endcsname{0.079}
\expandafter\def\csname result@additional-h4-axis-exterior-maximum_absolute_error_mha\endcsname{0.138}
\expandafter\def\csname result@additional-h4-axis-interior-mae_mha\endcsname{0.045}
\expandafter\def\csname result@additional-h4-axis-interior-maximum_absolute_error_mha\endcsname{0.092}
\expandafter\def\csname result@additional-h4-axis-mae_mha\endcsname{0.062}
\expandafter\def\csname result@additional-h4-axis-maximum_absolute_error_mha\endcsname{0.138}
\expandafter\def\csname result@additional-h4-grid-boundary-mae_mha\endcsname{0.132}
\expandafter\def\csname result@additional-h4-grid-boundary-maximum_absolute_error_mha\endcsname{0.186}
\expandafter\def\csname result@additional-h4-grid-exterior-mae_mha\endcsname{0.144}
\expandafter\def\csname result@additional-h4-grid-exterior-maximum_absolute_error_mha\endcsname{0.235}
\expandafter\def\csname result@additional-h4-grid-interior-mae_mha\endcsname{0.125}
\expandafter\def\csname result@additional-h4-grid-interior-maximum_absolute_error_mha\endcsname{0.157}
\expandafter\def\csname result@additional-h4-grid-mae_mha\endcsname{0.133}
\expandafter\def\csname result@additional-h4-grid-maximum_absolute_error_mha\endcsname{0.235}

\expandafter\def\csname result@cost-full-time-frozen-161\endcsname{0.845}
\expandafter\def\csname result@cost-full-time-frozen-41\endcsname{0.813}
\expandafter\def\csname result@cost-full-time-independent-161\endcsname{42.352}
\expandafter\def\csname result@cost-full-time-independent-41\endcsname{10.785}
\expandafter\def\csname result@cost-full-time-shared-161\endcsname{3.746}
\expandafter\def\csname result@cost-full-time-shared-41\endcsname{1.265}
\expandafter\def\csname result@cost-full-time-warm-161\endcsname{1.950}
\expandafter\def\csname result@cost-full-time-warm-41\endcsname{0.664}
\expandafter\def\csname result@cost-gpu-frozen-161\endcsname{1.644}
\expandafter\def\csname result@cost-gpu-frozen-41\endcsname{1.613}
\expandafter\def\csname result@cost-gpu-independent-161\endcsname{42.351}
\expandafter\def\csname result@cost-gpu-independent-41\endcsname{10.785}
\expandafter\def\csname result@cost-gpu-shared-161\endcsname{4.162}
\expandafter\def\csname result@cost-gpu-shared-41\endcsname{1.682}
\expandafter\def\csname result@cost-gpu-warm-161\endcsname{1.948}
\expandafter\def\csname result@cost-gpu-warm-41\endcsname{0.664}
\expandafter\def\csname result@ec-li2se-0.0001-energy_se_max_mha\endcsname{0.356}
\expandafter\def\csname result@ec-li2se-0.0001-energy_se_min_mha\endcsname{0.124}
\expandafter\def\csname result@ec-li2se-0.0001-mae_mha\endcsname{0.948}
\expandafter\def\csname result@ec-li2se-0.0001-max_absolute_error_mha\endcsname{1.073}
\expandafter\def\csname result@ec-li2se-0.001-energy_se_max_mha\endcsname{0.223}
\expandafter\def\csname result@ec-li2se-0.001-energy_se_min_mha\endcsname{0.105}
\expandafter\def\csname result@ec-li2se-0.001-mae_mha\endcsname{0.867}
\expandafter\def\csname result@ec-li2se-0.001-max_absolute_error_mha\endcsname{0.894}
\expandafter\def\csname result@ec-li2se-0.01-energy_se_max_mha\endcsname{0.114}
\expandafter\def\csname result@ec-li2se-0.01-energy_se_min_mha\endcsname{0.103}
\expandafter\def\csname result@ec-li2se-0.01-mae_mha\endcsname{4.399}
\expandafter\def\csname result@ec-li2se-0.01-max_absolute_error_mha\endcsname{10.265}
\expandafter\def\csname result@ec-li2se-1e-06-energy_se_max_mha\endcsname{0.356}
\expandafter\def\csname result@ec-li2se-1e-06-energy_se_min_mha\endcsname{0.124}
\expandafter\def\csname result@ec-li2se-1e-06-mae_mha\endcsname{0.948}
\expandafter\def\csname result@ec-li2se-1e-06-max_absolute_error_mha\endcsname{1.073}
\expandafter\def\csname result@ec-li2se-1e-08-energy_se_max_mha\endcsname{0.356}
\expandafter\def\csname result@ec-li2se-1e-08-energy_se_min_mha\endcsname{0.124}
\expandafter\def\csname result@ec-li2se-1e-08-mae_mha\endcsname{0.948}
\expandafter\def\csname result@ec-li2se-1e-08-max_absolute_error_mha\endcsname{1.073}
\expandafter\def\csname result@ec-li2se-1e-10-energy_se_max_mha\endcsname{0.356}
\expandafter\def\csname result@ec-li2se-1e-10-energy_se_min_mha\endcsname{0.124}
\expandafter\def\csname result@ec-li2se-1e-10-mae_mha\endcsname{0.948}
\expandafter\def\csname result@ec-li2se-1e-10-max_absolute_error_mha\endcsname{1.073}
\expandafter\def\csname result@ec-li2se-direct-energy_se_max_mha\endcsname{0.025}
\expandafter\def\csname result@ec-li2se-direct-energy_se_min_mha\endcsname{0.004}
\expandafter\def\csname result@ec-li2se-direct-mae_mha\endcsname{0.840}
\expandafter\def\csname result@ec-li2se-direct-max_absolute_error_mha\endcsname{1.048}
\expandafter\def\csname result@ec-li2se-direct-query-hours\endcsname{0.799}
\expandafter\def\csname result@ec-li2se-production-setup-hours\endcsname{2.669}
\expandafter\def\csname result@ec-li2se-query-seconds\endcsname{0.039}

\expandafter\def\csname result@fig2-n2-anchors-count\endcsname{12}
\expandafter\def\csname result@fig2-n2-anchors-mae_mha\endcsname{0.243}
\expandafter\def\csname result@fig2-n2-anchors-maximum_absolute_error_mha\endcsname{0.397}
\expandafter\def\csname result@fig2-n2-anchors-within_1p6_count\endcsname{12}
\expandafter\def\csname result@fig2-n2-queries-count\endcsname{69}
\expandafter\def\csname result@fig2-n2-queries-mae_mha\endcsname{0.274}
\expandafter\def\csname result@fig2-n2-queries-maximum_absolute_error_mha\endcsname{1.025}
\expandafter\def\csname result@fig2-n2-queries-within_1p6_count\endcsname{69}
\expandafter\def\csname result@fig2-co-anchors-count\endcsname{11}
\expandafter\def\csname result@fig2-co-anchors-mae_mha\endcsname{0.052}
\expandafter\def\csname result@fig2-co-anchors-maximum_absolute_error_mha\endcsname{0.079}
\expandafter\def\csname result@fig2-co-anchors-within_1p6_count\endcsname{11}
\expandafter\def\csname result@fig2-co-queries-count\endcsname{60}
\expandafter\def\csname result@fig2-co-queries-mae_mha\endcsname{0.173}
\expandafter\def\csname result@fig2-co-queries-maximum_absolute_error_mha\endcsname{1.059}
\expandafter\def\csname result@fig2-co-queries-within_1p6_count\endcsname{60}
\expandafter\def\csname result@fig2-li2se-anchors-count\endcsname{5}
\expandafter\def\csname result@fig2-li2se-anchors-mae_mha\endcsname{0.820}
\expandafter\def\csname result@fig2-li2se-anchors-maximum_absolute_error_mha\endcsname{0.875}
\expandafter\def\csname result@fig2-li2se-anchors-within_1p6_count\endcsname{5}
\expandafter\def\csname result@fig2-li2se-queries-count\endcsname{16}
\expandafter\def\csname result@fig2-li2se-queries-mae_mha\endcsname{0.831}
\expandafter\def\csname result@fig2-li2se-queries-maximum_absolute_error_mha\endcsname{0.888}
\expandafter\def\csname result@fig2-li2se-queries-within_1p6_count\endcsname{16}

\title{Foundation Neural-Network Quantum States\\for Molecular Potential Energy Surfaces\\in Second Quantization}
\iflatexml\else
  \qkpdftitle{Foundation Neural-Network Quantum States for Molecular Potential Energy Surfaces in Second Quantization}
\fi
\iflatexml
  \author[1]{Lizhong Fu}
  \author[2,3]{Jianan Wei}
  \author[2,3,*]{Wenguan Wang}
  \author[1,*]{Honghui Shang}
  \affil[1]{State Key Laboratory of Precision and Intelligent Chemistry, Hefei National Laboratory, University of Science and Technology of China}
  \affil[2]{State Key Lab of Brain-Machine Intelligence, Zhejiang University}
  \affil[3]{College of Artificial Intelligence, Zhejiang University}
  \affil[*]{Corresponding authors}
\else
  \author{Lizhong Fu\textsuperscript{1}\quad Jianan Wei\textsuperscript{2,3}\quad Wenguan Wang\textsuperscript{2,3,*}\quad Honghui Shang\textsuperscript{1,*}}
  \qkaffiliations{\textsuperscript{1}State Key Laboratory of Precision and Intelligent Chemistry, Hefei National Laboratory,\\
  \hspace*{0.5em}University of Science and Technology of China\\
  \textsuperscript{2}State Key Lab of Brain-Machine Intelligence, Zhejiang University\\
  \textsuperscript{3}College of Artificial Intelligence, Zhejiang University}
  \qkauthornote{* Corresponding authors}
  \qkpdfauthor{Lizhong Fu; Jianan Wei; Wenguan Wang; Honghui Shang}
\fi

\date{27 November 2026}
\begin{document}
\begin{qkfrontmatter}
\maketitle
\begin{abstract}
Second-quantized neural-network quantum states have achieved accurate molecular energies, but extending them across molecular geometries requires a shared representation of the geometry-dependent wavefunction coefficients. We introduce geometry-conditioned foundation neural-network quantum states for molecular electronic structure in second quantization. A single autoregressive model learns a family of ground states from sparse anchor geometries and provides wavefunctions at untrained geometries without further optimization. Orbital alignment matches orbital identities and transports their phases, establishing an aligned orbital basis across geometries. Frozen energies reach chemical accuracy at every query in the primary N$_2$, CO, H$_4$, and Li$_2$Se benchmarks. On additional molecular paths, the energy-trained wavefunctions yield dipoles, quadrupoles, and natural occupations without property labels. Across three paired N$_2$ training seeds, orbital alignment lowers the mean absolute energy error over all untrained query geometries from 34--37~mHa to 0.049--0.085~mHa. At approximately 1~mHa mean absolute error, frozen evaluation reduces the per-geometry cost by $986\times$ relative to independent optimization, yielding an estimated $25.8\times$ end-to-end GPU-cost reduction on a 161-point N$_2$ grid.

\end{abstract}
\qkkeywords{Neural-network quantum states, potential energy surfaces, second quantization, variational Monte Carlo}
\end{qkfrontmatter}
\section{Introduction}
Molecular potential energy surfaces (PESs) connect electronic structure to vibrations, conformational changes, and chemical reactions. Neural-network quantum states (NQSs) provide an expressive ansatz for many-body wavefunctions \citep{carleo2017,hermann2023review}. For molecules in second quantization, NQSs parameterize the wavefunction coefficients in an occupation basis and achieve accurate electronic energies through \textit{ab initio} variational optimization \citep{choo2020,barrett2022,zhao2023chem,liu2024backflow,kan2025bridge,shang2025}. Constructing a PES with independently optimized NQSs, however, repeats this optimization at each geometry, leaving the shared electronic structure along the molecular path largely unused \citep{wu2025pes}.

Sharing an NQS across a family of Hamiltonians offers a route to amortizing repeated wavefunction optimization. Foundation neural-network quantum states (FNQSs) establish this principle for quantum spin systems by conditioning a shared model on Hamiltonian couplings \citep{rende2025fnqs}. Extending this framework to molecular electronic structure in second quantization requires a compact conditioning representation for the molecular Hamiltonian family. In this work, we introduce a geometry-conditioned FNQS that combines molecular geometry with orbital descriptors to provide global and orbital-specific conditioning. Jointly trained at sparse anchor geometries along a PES, the model provides wavefunctions at untrained geometries with frozen parameters, from which energies and electronic observables are evaluated.

Sharing a second-quantized wavefunction model across molecular geometries also requires a consistent orbital basis \citep{bensberg2023,rath2025}. Independently generated molecular orbitals can change order and sign across geometries. The same occupation string can then refer to different orbital identities or phase conventions, introducing artificial discontinuities in the wavefunction coefficients even when the physical state varies smoothly. We align orbital identities and phases along the path to establish a consistent basis for the shared model. Ablation studies show that this orbital consistency enables accurate predictions at untrained geometries.

Our contributions are:
\begin{enumerate}
\item \textbf{FNQSs for molecular second quantization.} To our knowledge, this is the first variationally trained, geometry-conditioned NQS for molecular electronic structure in second quantization that predicts wavefunctions at untrained geometries with frozen parameters. The primary N$_2$, CO, H$_4$, and Li$_2$Se query sets all reach chemical accuracy. Energy-trained wavefunctions also yield dipoles, quadrupoles, and natural occupations without property-specific training.
\item \textbf{Orbital consistency for accurate parameter sharing.} We combine orbital matching and phase alignment with geometry and per-orbital conditioning. Across three paired N$_2$ seeds, alignment reduces the mean absolute errors over all non-anchor queries from 34--37 to 0.049--0.085~mHa, demonstrating that orbital consistency enables accurate frozen transfer.
\item \textbf{Amortized computation without query-time optimization.} On a 161-point N$_2$ grid, our method yields estimated end-to-end GPU-cost reductions of $25.8\times$ over independent optimization and $2.53\times$ over shared pretraining with fine-tuning, using fixed budgets with MAEs near 1~mHa. The comparison includes initial training and energy evaluation; a cost model separates this initial investment from the cost of additional geometries.
\end{enumerate}

\section{Related work}\label{sec:related}
\paragraph{Shared real-space wavefunctions and potential energy surfaces.}
PESNet trains a single real-space wavefunction across molecular geometries, reducing repeated variational optimization \citep{gao2022}. PlaNet adds an energy surrogate trained on intermediate VMC estimates, enabling energy queries without Monte Carlo sampling at inference \citep{gao2023}. Both methods use real-space formulations, parameterizing electronic wavefunctions over continuous electron coordinates \citep{pfau2020,hermann2020,vonglehn2023}. Related work extends reuse across molecules through transferable orbital constructions and pretrained wavefunctions \citep{scherbela2024,gao2023globe,foster2025}. We instead adopt a second-quantized autoregressive representation, which facilitates comparison with conventional quantum-chemical methods and evaluation of electronic observables, while also enabling direct sampling under particle-number and symmetry constraints \citep{barrett2022,zhao2023chem,malyshev2023sym}.

\paragraph{Foundation neural-network quantum states.}
FNQSs condition a shared spin wavefunction on Hamiltonian couplings \citep{rende2025fnqs}. Their embedding concatenates $O(1)$ global couplings to each configuration patch, or pairs patches of $O(N)$ couplings with the corresponding configuration patches, where $N$ is the number of spins. The authors identify second-quantized fermions and molecular systems as directions for extension. Molecular two-electron integrals carry four orbital indices and number $O(M^4)$ for $M$ orbitals, so their full tensor does not directly fit either input construction. Molecular conditioning therefore calls for an appropriate encoding of the Hamiltonian family, alongside orbital alignment to maintain a consistent meaning for the learned coefficients. We use geometry and aligned per-orbital descriptors for this encoding.

\paragraph{Wavefunction interpolation.}
Eigenvector continuation (EC) interpolates wavefunctions by solving the query Hamiltonian in the span of anchor states \citep{frame2018,mejuto2023}. \citet{rath2025} transfer anchor coefficients in consistently indexed, symmetrically orthogonalized atomic orbitals. We align HF molecular orbitals to retain their concentrated occupation distributions for autoregressive sampling. Applying EC to sampled NQSs requires estimating cross-state overlaps and Hamiltonian matrix elements. Near-linear dependence among anchor states can amplify sampling errors in the resulting generalized eigenvalue problem and produce spurious low-energy solutions; overlap-spectrum truncation and other noise-aware regularization methods address this sensitivity \citep{epperly2022qsd,hicks2023trimmed}. We instead learn a geometry-conditioned wavefunction whose frozen-query evaluation requires neither cross-state matrix elements nor a generalized eigenvalue solve.

\begin{figure*}[t]
\centering
\includegraphics[width=\linewidth]{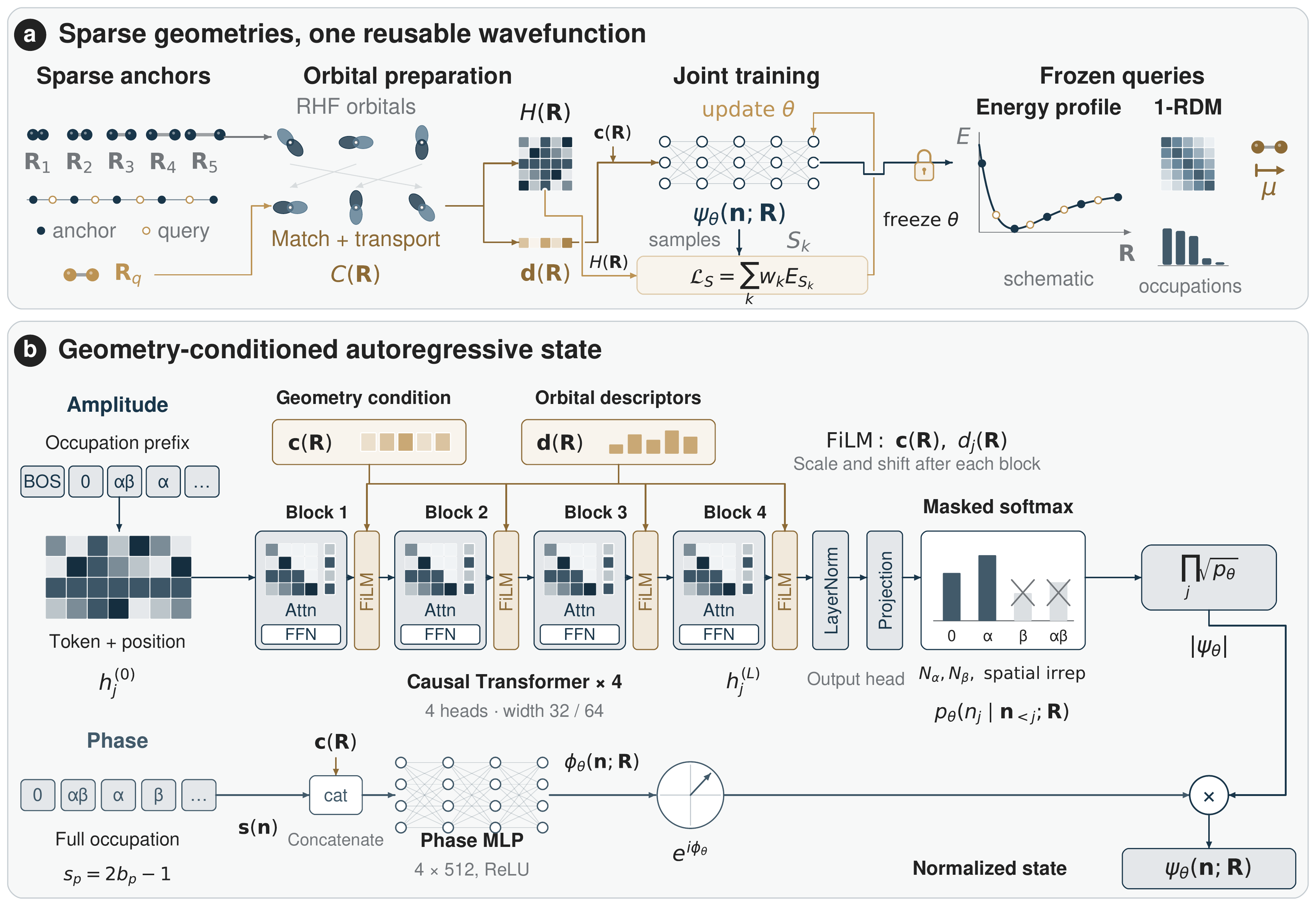}
\caption{\textbf{Geometry-conditioned FNQSs.} (a) Starting from sparse anchor geometries, restricted Hartree--Fock (RHF) orbitals are matched and phase-aligned to construct consistent occupation bases and molecular Hamiltonians. Joint variational Monte Carlo (VMC) training produces a shared wavefunction model. At new geometries, the frozen model yields energies and the one-particle reduced density matrix (1-RDM). (b) Occupation prefixes pass through causal Transformer blocks, with geometry and orbital descriptors supplied through feature-wise linear modulation (FiLM), to produce normalized occupation probabilities and hence the wavefunction amplitude. A separate multilayer perceptron (MLP) maps the full occupation string and geometry to a phase. Amplitude and phase combine into the wavefunction.}
\label{fig:method}
\end{figure*}

\section{Method}\label{sec:method}
\paragraph{Problem setup.}\label{sec:setup}
Fix a molecule, an atomic-orbital basis set, and spin populations $(N_\alpha,N_\beta)$ in $M$ retained spatial orbitals. The electronic Hamiltonian at nuclear geometry $\R$ is
\begin{equation}
\begin{aligned}
H(\R)={}&E_{\mathrm c}(\R)+\sum_{pq}h_{pq}(\R)a_p^\dagger a_q\\
&+\frac12\sum_{pqrs}v_{pqrs}(\R)a_p^\dagger a_q^\dagger a_s a_r,
\end{aligned}
\label{eq:hamiltonian}
\end{equation}
where $p,q,r,s$ label spin orbitals, $h_{pq}$ and $v_{pqrs}=\langle pq|rs\rangle$ are the one- and two-electron integrals in physicists' notation and $E_{\mathrm c}$ collects nuclear repulsion and frozen-core terms. A state is a coefficient function $\psi(\nn;\R)$ over occupation strings $\nn=(n_1,\ldots,n_M)$ with $n_j\in\{0,\alpha,\beta,\alpha\beta\}$. The task is to build a shared model for the ground-state wavefunction of a family $\{H(\R)\}$ along a molecular path.

\paragraph{Overview.} Fig.~\ref{fig:method} summarizes the construction. We first align orbital bases across geometries (Section~\ref{sec:coords}), then condition an autoregressive wavefunction on geometry and aligned orbital descriptors (Section~\ref{sec:conditioning}). Joint variational optimization at sparse anchors learns shared parameters, which remain frozen when evaluating energies and electronic observables at new geometries (Section~\ref{sec:training}).

\subsection{Orbital alignment across geometries}\label{sec:coords}
The choice of orbital basis affects the expressivity of a restricted variational ansatz \citep{moreno2023}. The occupation basis is built from RHF orbitals. In this basis, molecular wavefunctions often concentrate much of their probability mass on a relatively small subset of configurations \citep{malyshev2024}. Batched autoregressive sampling (BAS) exploits this concentration by aggregating repeated configurations \citep{barrett2022}.

Sharing coefficients across geometries requires a consistent convention for these orbitals. Sorting orbitals by energy at each geometry does not ensure this consistency: independent RHF calculations can return orbitals with arbitrary signs, and energy crossings can change their order. A sign flip of spin-orbital $p$ multiplies $\psi(\nn;\R)$ by $-1$ if $p$ is occupied, and an orbital permutation relabels configurations and transforms their coefficients with the corresponding fermionic parity. These changes in the orbital basis can make the coefficients discontinuous even when the physical state varies smoothly, complicating learning with a continuous geometry-conditioned model.

Learning from eigenvectors requires accounting for their nonunique representation: SignNet and BasisNet address sign and degenerate-basis ambiguities in spectral graph learning \citep{lim2023sign}, while transferable neural orbitals have incorporated orbital-sign equivariance \citep{scherbela2024}. In an occupation-basis model, orbital transformations act on the wavefunction coefficients themselves. We handle this dependence by aligning orbital identities and phases before sharing the autoregressive model across geometries.

For orbital coefficient matrices $C_A,C_B$ at neighboring geometries, the overlap is $O_{AB}=C_A^\dagger S_{AB}C_B$, where $S_{AB}$ contains the overlaps between their atomic-orbital (AO) bases. Within compatible occupied/virtual, frozen-core, and symmetry classes, maximum-overlap assignment \citep{kuhn1955} selects the permutation $\pi^*=\arg\max_\pi\sum_j |(O_{AB})_{j,\pi(j)}|$.
The target orbitals are reordered by $\pi^*$ and phase-aligned to make matched overlaps real and nonnegative. Where exact degeneracies leave symmetry labels ambiguous, orbitals are first resolved with respect to a fixed physical symmetry operation, as in the NH$_3$ construction detailed in Appendix~\ref{app:coordinates}. One- and two-electron integrals are transformed into the aligned orbital basis, in which orbital descriptors are also evaluated.

\subsection{Geometry-conditioned autoregressive wavefunctions}\label{sec:conditioning}
Using aligned orbitals with coefficient matrix $C(\R)$, the model represents the state as
\begin{equation}
\begin{aligned}
\ket{\Psi_\theta(\R)}
&=\sum_{\nn\in\Omega}\psi_\theta(\nn;\R)\ket{\nn;C(\R)},\\
\psi_\theta(\nn;\R)
&=e^{i\phi_\theta(\nn;\R)}\prod_{j=1}^{M}\sqrt{p_\theta(n_j\mid \nn_{<j};\R)}.
\end{aligned}
\label{eq:ansatz}
\end{equation}
Here $\Omega$ contains occupation strings with the prescribed spin populations and, when imposed, total spatial irreducible representation (irrep), and $\nn_{<j}$ denotes the occupations preceding orbital $j$. Before each conditional is normalized, a mask excludes choices that cannot be completed to a configuration in $\Omega$, using an exact backward reachability table when spatial symmetry is imposed. The resulting conditionals define a normalized state and permit direct autoregressive sampling \citep{sharir2020,hibat2020,barrett2022,malyshev2023sym}. The amplitude network uses $L$ causal Transformer blocks, each combining attention and a feed-forward network (FFN), to predict each occupation from its prefix \citep{vaswani2017,shang2025}. The shifted token and position embeddings in Fig.~\ref{fig:method}(b) are $h_j^{(0)}=e_{\mathrm{tok}}(n_{j-1})+e_{\mathrm{pos}}(j)$, where $n_0$ is the beginning-of-sequence (BOS) token.

For a fixed molecule, basis set, and electronic sector, geometry parameterizes the Hamiltonian family along the chosen path. We therefore use geometry as a compact global condition, supplemented by aligned orbital descriptors that provide orbital-specific electronic information at each position in the occupation sequence. Geometry enters through a global condition $\mathbf c(\R)$, such as a bond length or a normalized path coordinate, and a vector $\mathbf d(\R)$ with one orbital descriptor per aligned orbital. Each component is the aligned Fock diagonal relative to the mean over active occupied orbitals, scaled by one hartree (Ha): $d_j(\R)=\bigl(F^{\mathrm{aligned}}_{jj}(\R)-\overline F_{\mathrm{occ}}(\R)\bigr)/(1\,\mathrm{Ha})$, where $F^{\mathrm{aligned}}$ is the Fock matrix in the aligned orbital basis and $\overline F_{\mathrm{occ}}$ is that mean. Under a signed permutation, its diagonal entries equal the canonical orbital energies.

After each Transformer block, a FiLM adapter \citep{perez2018film} modulates the hidden features using both inputs. For block output $u^{(\ell)}=\mathrm{Transformer}_{\ell}(h^{(\ell-1)})$, the FiLM update is
\begin{equation}
h_j^{(\ell)}=\bigl(1+\gamma_{\ell}^{c}(\mathbf c)+\gamma_{\ell}^{d}(d_j)\bigr)\odot u_j^{(\ell)}+\beta_{\ell}^{c}(\mathbf c)+\beta_{\ell}^{d}(d_j).
\label{eq:film}
\end{equation}
The $\gamma$ and $\beta$ maps produce feature-wise scales and shifts, and $\odot$ denotes element-wise multiplication. The geometry maps are affine, while the descriptor maps are linear. All four maps are initialized to zero, so each adapter initially acts as the identity. After the final block, layer normalization (LayerNorm) \citep{ba2016layer} and a linear projection produce logits for the masked softmax, yielding $p_\theta(n_j\mid \nn_{<j};\R)$ (Fig.~\ref{fig:method}(b)). Implementation and optimizer settings are given in Appendix~\ref{app:sampling}.

A separate phase MLP with rectified linear unit (ReLU) activations takes the concatenation of $\mathbf c(\R)$ and the full occupation string encoded as $\mathbf s(\nn)\in\{-1,1\}^{2M}$, with $s_p=2b_p-1$ for spin-orbital occupation $b_p\in\{0,1\}$. It predicts the angle $\phi_\theta(\nn;\R)$, whose factor $e^{i\phi_\theta}$ combines with the autoregressive amplitude in Equation~\eqref{eq:ansatz}. Orbital descriptors enter only the amplitude branch.

\subsection{Joint variational training and frozen inference}\label{sec:training}
The model is trained using a first-principles VMC method. The geometry-dependent molecular Hamiltonians supply the variational objective directly, without labeled energies or properties. At each optimization update, BAS propagates counts through successive occupation prefixes, drawing configurations at each anchor geometry from the corresponding wavefunction distribution \citep{barrett2022}. A joint energy estimator constructed from these samples \citep{wu2023sample,malyshev2024} is minimized. Let $S_k$ contain the distinct configurations sampled at anchor $k$, $\psi_{S_k}$ their coefficient vector, and $H_{S_kS_k}$ the corresponding Hamiltonian submatrix. The joint objective is
\begin{equation}
\begin{aligned}
E_{S_k}(\theta;\R_k)
&=\frac{\psi_{S_k}^\dagger H_{S_kS_k}(\R_k)\psi_{S_k}}
{\psi_{S_k}^\dagger\psi_{S_k}},\\
\mathcal L_S(\theta)
&=\sum_k w_k E_{S_k}(\theta;\R_k),
\end{aligned}
\label{eq:projected}
\end{equation}
where $w_k$ are the prescribed anchor weights. All anchors contribute to every joint update, extending the shared-training infrastructure of \citet{wu2025pes}. Appendix~\ref{app:sampling} gives the explicit gradient and sampling implementation. The energy of the complete network state is evaluated separately using the frozen-query procedure below.

\paragraph{Frozen queries and electronic observables.}\label{sec:inference}
At a query geometry $\R_q$, orbital preparation supplies the aligned Hamiltonian and the geometry and orbital inputs (Fig.~\ref{fig:method}(a)). The network parameters remain frozen while fresh configurations are drawn from $|\psi_\theta(\nn;\R_q)|^2$. For each sampled configuration, the local energy includes all determinants connected by the Hamiltonian:
\begin{equation}
\begin{aligned}
E_{\mathrm{loc},\theta}(\nn;\R)
&=\sum_{\nn'}H_{\nn\nn'}(\R)
\frac{\psi_\theta(\nn';\R)}{\psi_\theta(\nn;\R)},\\
E_\theta(\R)
&=\E_{|\psi_\theta|^2}\operatorname{Re}E_{\mathrm{loc},\theta}.
\end{aligned}
\label{eq:local}
\end{equation}
The expectation is taken over the normalized network distribution at the query geometry. It is a variational upper bound to the ground-state energy in the chosen orbital space and sector; its Monte Carlo estimate carries sampling uncertainty, which is reported.

Replacing $H$ by another operator gives its expectation in the same wavefunction. This yields the spin-summed 1-RDM, $\gamma_{ij}=\sum_\sigma\langle a_{i\sigma}^\dagger a_{j\sigma}\rangle$, where $i,j$ label spatial orbitals. Its eigenvalues are the natural occupations. Molecular dipoles are obtained from the density matrix, including the frozen-core and nuclear contributions.

\section{Experiments}\label{sec:exp}

\begin{figure*}[t]
\centering\includegraphics[width=\linewidth]{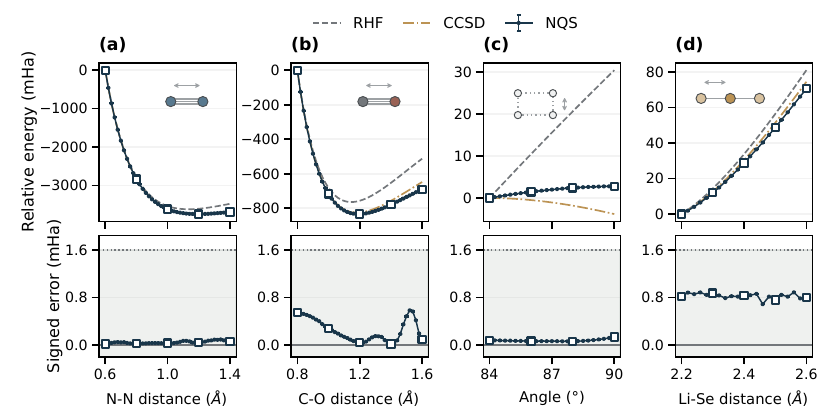}
\caption{\textbf{Potential energy surfaces.} (a) N$_2$, (b) CO, (c) H$_4$, and (d) Li$_2$Se. Each panel combines NQS, RHF, and coupled-cluster singles and doubles (CCSD) relative energies (upper) and NQS signed errors in the lower panels, with the 1.6~mHa band. Error references are FCI in (a--c) and DMRG in (d). Energies are relative to each method's path origin. CCSD uses a restricted reference. Squares mark anchors, dots queries, and bars one sampling standard error.}
\label{fig:representative-pes}
\end{figure*}

\begin{figure*}[t]
\centering\includegraphics[width=\linewidth]{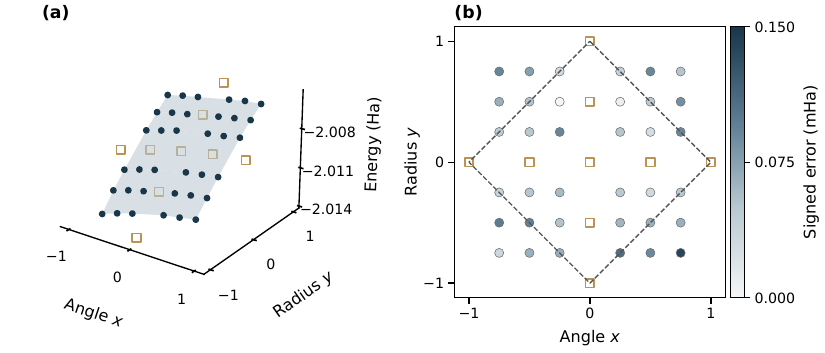}
\caption{\textbf{Two-dimensional H$_4$ transfer.} (a) Absolute NQS energies at the 36 frozen queries combining angular and radial distortions; the surface connects the evaluated grid points. (b) Signed energy errors against FCI at the same queries. Hollow squares in both panels mark axis-only training anchors; the dashed diamond in (b) is their convex hull. Coordinates are normalized as specified in Appendix~\ref{app:h4-composition}.}
\label{fig:geometric-transfer}
\end{figure*}

\begin{figure*}[t]
\centering\includegraphics[width=\linewidth]{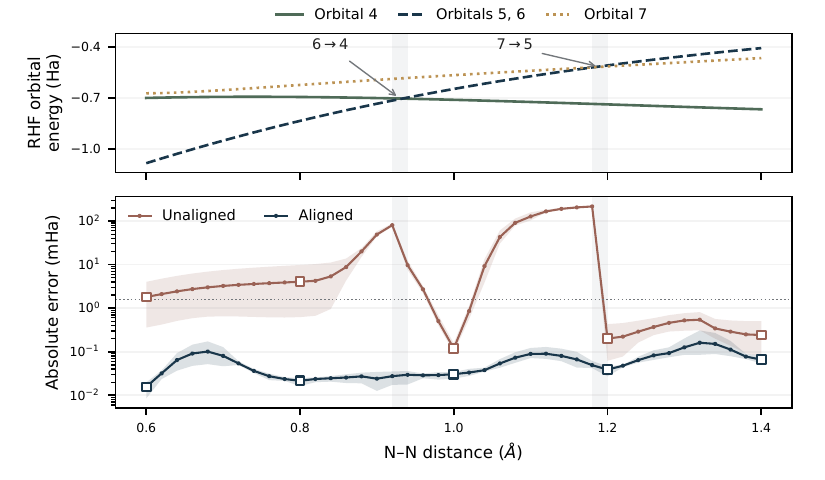}
\caption{\textbf{Orbital alignment.} Alignment enables transfer between N$_2$ training anchors. The upper plot tracks aligned RHF orbital energies; the degenerate pair 5/6 is drawn as one curve. Arrows label changes in the canonical indices of the indicated continuous branches. Both plots share the N--N distance axis; vertical bands mark the two orbital-ordering exchanges (0.92--0.94 and 1.18--1.20~\AA). The lower plot compares absolute energy errors over three paired seeds: lines show means and colored ribbons the minimum--maximum range. Squares mark anchors, dots queries, and the dotted line 1.6~mHa. Architecture, optimization, and sampling budgets are matched within each pair.}
\label{fig:component}
\end{figure*}

\begin{figure*}[t]
\centering\includegraphics[width=\linewidth]{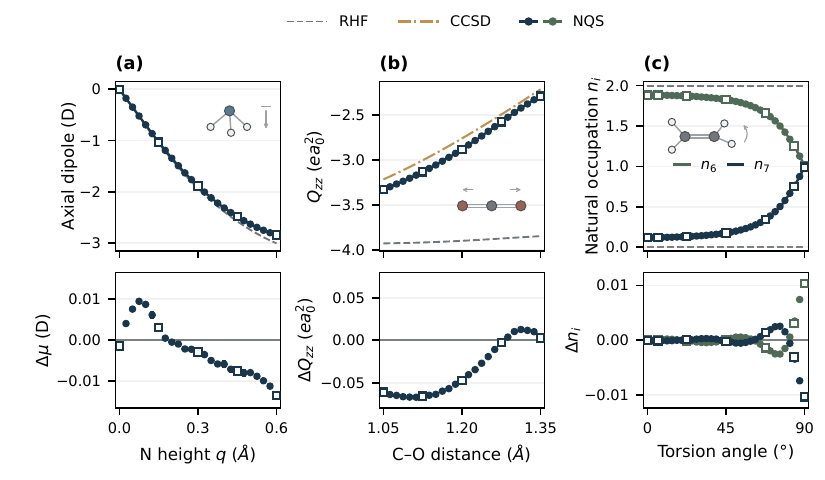}
\caption{\textbf{Electronic properties.} (a) NH$_3$ dipole, (b) CO$_2$ quadrupole, and (c) ethylene natural occupations 6/7. Each panel combines the observable (upper) and its NQS-minus-FCI error (lower). Upper curves show NQS estimates connected by solid lines, RHF (dashed), and CCSD (dash-dotted). Circles mark queries, squares anchors, and bars one sampling standard error.}
\label{fig:molecular-properties}
\end{figure*}

\begin{figure*}[t]
\centering\includegraphics[width=\linewidth]{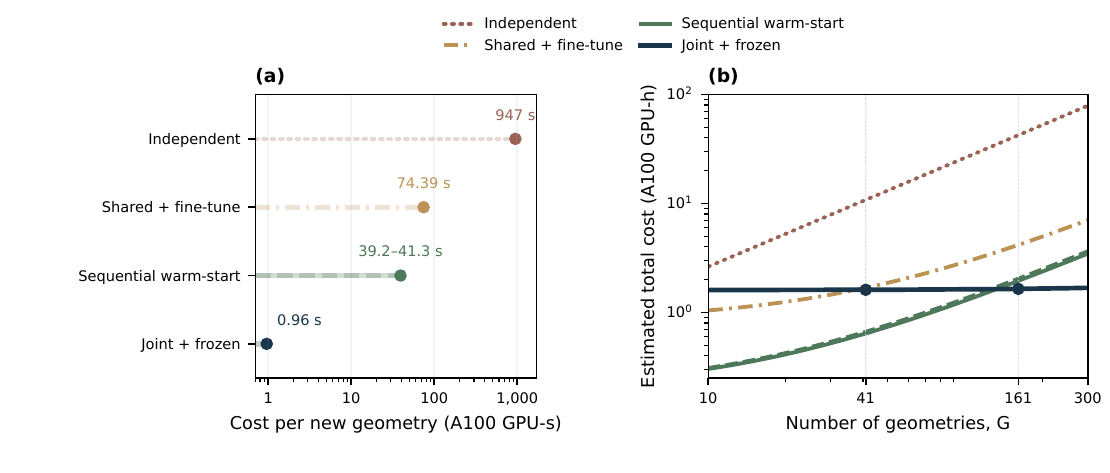}
\caption{\textbf{Computational cost.} The N$_2$ workflows use fixed training budgets with MAEs near 1~mHa. (a) Average cost per additional geometry, including optimization where applicable and energy evaluation. The warm-start range spans the measured means at 41- and 161-point spacings. (b) Estimated total costs, including initial training, on logarithmic axes. Dots and vertical guides mark frozen evaluation on the 41- and 161-point grids. Independent and shared-adaptation costs are estimated from eight geometries selected by stratified sampling; warm-start dashed and solid curves use the 41- and 161-point spacings, respectively. Each curve holds its measured average per-geometry cost fixed as the query count varies.}
\label{fig:cost-scaling}
\end{figure*}

\paragraph{Systems and evaluation.}\label{sec:benchmark}
We use STO-3G \citep{hehre1969sto} for the N$_2$, CO, and all-electron Li$_2$Se bond scans, with the Se basis from \citet{pietro1980sto}. The H$_4$ rectangular deformation uses cc-pVDZ \citep{dunning1989}. For NH$_3$ inversion we use 6-31G, with H functions from \citet{ditchfield1971basis} and N functions from \citet{hehre1972basis}. The CO$_2$ symmetric stretch and C$_2$H$_4$ torsion use STO-3G. For each system, a model is trained jointly at sparse anchors and evaluated with frozen parameters at intervening query geometries. Appendix~\ref{app:protocols} specifies the paths, geometry splits, and retained electronic spaces; Appendix~\ref{app:sampling} gives the network and optimization settings.

Full configuration interaction (FCI) calculations provide the reference values in the same basis and frozen-core space for all systems except Li$_2$Se, following the validation protocol for second-quantized neural states \citep{choo2020,barrett2022,shang2025}. Li$_2$Se uses density matrix renormalization group (DMRG) references (Appendix~\ref{app:li2se}). For an evaluation set $Q$, mean absolute error (MAE) is $|Q|^{-1}\sum_{q\in Q}|E_\theta(\R_q)-E_{\mathrm{ref}}(\R_q)|$; we also report the maximum absolute error over $Q$ and use 1.6~mHa as the chemical-accuracy threshold. Anchor and query errors are kept separate. The main energy benchmarks report statistics over all non-anchor queries. Neural energies use Equation~\eqref{eq:local}, evaluated using three independent sampling replicates per geometry. Sampling error bars give one standard error of the mean. Property errors use the corresponding FCI observables at the same queries.

\subsection{Frozen energy predictions across molecular paths}\label{sec:frozen-main}
Sparse-anchor training yields accurate energy profiles at untrained geometries. In Fig.~\ref{fig:representative-pes}, a single frozen model per system resolves the energy profiles of N$_2$, CO, and H$_4$, with every anchor and query within 1.6~mHa of FCI. N$_2$ and CO query MAEs of \result{final-n2-query-mae} and \result{final-co-query-mae}~mHa show that anchor accuracy carries across both homonuclear and heteronuclear bond scans without further optimization. Appendix~\ref{app:expanded-coverage} extends both bond scans to 2.20~\AA{} with additional training anchors.

H$_4$ tests transfer on a finer energy scale: its rise toward the square geometry spans only a few mHa. The model resolves this shallow multireference profile in the larger cc-pVDZ orbital space with a maximum query error of \result{final-h4-query-max}~mHa. Thus, the shared wavefunction captures both broad bond-stretching curves and small energy differences along deformation. For Li$_2$Se (40 electrons in 28 orbitals), the five-anchor model gives a query MAE of \result{fig2-li2se-queries-mae_mha}~mHa and a maximum error of \result{fig2-li2se-queries-maximum_absolute_error_mha}~mHa against DMRG over all 16 final queries. Appendix~\ref{sec:frozen} gives the full anchor and query statistics.

\paragraph{Transfer to unseen combinations.}
H$_4$ tests whether the shared state combines geometric changes seen only separately during training. Nine anchors vary either the angle or the radius, while all 36 frozen queries combine both distortions. The model resolves the resulting two-dimensional energy surface with MAE \result{additional-h4-axis-mae_mha}~mHa and maximum error \result{additional-h4-axis-maximum_absolute_error_mha}~mHa, including 12 queries outside the training convex hull (Fig.~\ref{fig:geometric-transfer}; Appendix~\ref{app:h4-composition}).

\subsection{Orbital alignment enables accurate transfer}\label{sec:mechanism-main}

Fig.~\ref{fig:component} compares the errors with unaligned and aligned orbitals along the N$_2$ scan across three paired training seeds. Within each pair, only the orbital basis changes; the architecture, optimizer, and sampling budget are identical. The aligned models predict every untrained geometry within chemical accuracy (query MAE 0.049--0.085~mHa across seeds). The unaligned models fit most anchors but fail between them, with query MAEs of 34--37~mHa and errors above 100~mHa adjacent to anchors; some unaligned runs also miss chemical accuracy at an anchor. The largest failures occur where canonical orbital ordering changes (upper plot in Fig.~\ref{fig:component}). Alignment removes these artificial discontinuities in the wavefunction coefficients, allowing accuracy at the anchors to extend to the intervening geometries.

Within the aligned orbital basis, conditioning ablations show smaller gains from FiLM and orbital descriptors (Appendix~\ref{sec:mechanism}).

\subsection{Electronic observables from the shared wavefunction}\label{sec:observables-main}
The model supplies a complete electronic wavefunction, giving access to observables from the same state used to evaluate energy. Fig.~\ref{fig:molecular-properties} probes charge displacement, spatial charge distribution, and frontier occupations. Each observable is computed from the one-particle reduced density matrix of the frozen, energy-trained state, without property labels or additional fitting.

The NH$_3$ dipole follows charge redistribution during inversion with query MAE \result{nh3-dipole-mae}~debye (D), comparable to CCSD. Symmetric CO$_2$ has no permanent dipole; its quadrupole instead reveals the changing charge distribution as the bonds lengthen. The model follows this variation with query MAE \result{co2-quadrupole-mae}~$e a_0^2$, below the CCSD error. Ethylene probes the growing multireference character during torsion: its frontier occupations approach one another toward the twisted geometry, a change that a restricted single determinant cannot describe. Their query MAE, averaged over both displayed occupations, is \result{c2h4-occupation-mae}, also below the CCSD error. These complementary observables show that the shared state retains useful electronic-structure information beyond its energy. The lower panels resolve signed errors; Appendix~\ref{sec:paths} reports energies from the same checkpoints and RHF/CCSD comparisons, and Appendix~\ref{app:protocols} defines the observables.

\subsection{Computational efficiency}\label{sec:baselines-main}
Sharing a wavefunction model reduces the work required at each new geometry. We quantify this benefit on the N$_2$ bond scan by comparing four workflows: independent optimization at every geometry, shared pretraining followed by per-geometry fine-tuning \citep{wu2025pes}, sequential warm-start optimization from the preceding geometry, and joint training of our geometry-conditioned FNQS followed by frozen evaluation. The shared-pretraining baseline uses the same NQS backbone and aligned orbitals but omits geometry conditioning. It minimizes the average loss over the training anchors before its weights are fine-tuned independently at each query.

The cost comparison uses separate fixed-budget checkpoints from the main accuracy benchmarks, targeting an MAE of approximately 1~mHa; Appendix~\ref{sec:baselines} reports the budgets, measured errors, and timing details. We include both initial training and subsequent queries in the total A100 GPU-hour cost. For each workflow $m$, the estimated cost of $G$ geometries takes the linear form
\begin{equation}
\widehat C_m(G)=a_m+G b_m.
\label{eq:surface-speedup}
\end{equation}
The intercept $a_m$ represents the fixed cost, and the slope $b_m$ the average cost of each additional geometry. Independent optimization has zero intercept: each geometry requires training from scratch and an energy estimate. Shared pretraining contributes a one-time cost, followed by fine-tuning and an energy estimate at each geometry. Sequential warm starts require training from scratch at the first geometry, followed by fine-tuning and an energy estimate at each subsequent geometry. Our geometry-conditioned FNQS incurs a one-time joint-training cost, after which each geometry requires only an energy estimate with frozen parameters (Fig.~\ref{fig:cost-scaling}).

Frozen evaluation reduces the cost of each additional geometry by approximately $986\times$ relative to independent optimization, $77\times$ relative to shared pretraining with fine-tuning, and $41$--$43\times$ relative to sequential warm starts. On the evaluated 161-point N$_2$ grid, the total cost is 1.644 GPU-hours, giving estimated end-to-end GPU-cost reductions of $25.8\times$, $2.53\times$, and $1.18\times$, respectively. Denser queries spread the initial training investment over more geometries, bringing the overall speedup closer to the per-geometry ratio. Sequential warm starts impose dependencies along the path; the other workflows permit parallel computation across geometries, including multi-geometry training for the shared and conditioned models. At the two measured warm-start rates, the savings from successive frozen queries cover the up-front cost of joint training at about 130 geometries. On the 161-point grid, the full-grid MAEs are 0.822~mHa for frozen evaluation and 0.993~mHa for warm starts.

RHF, orbital alignment, and integral preparation take 6.91~s for the entire 161-point grid, less than 0.3\% of the total elapsed time for either workflow (Appendix~\ref{sec:baselines}).

The four NNQS workflows in Fig.~\ref{fig:cost-scaling}, which varies the number of geometries on N$_2$, use sampling and can keep the same evaluation procedure as the determinant space grows. We compare EC separately because full enumeration is a natural choice for small systems and becomes impractical for larger systems. For N$_2$ and CO, all 14,400 coefficients of each anchor can be enumerated, so overlaps are direct inner products and transition densities can be computed once. With these densities cached, each query contraction and eigensolve takes 1--2~ms. Enumerating $9.66\times10^{12}$ coefficients per anchor is impractical on Li$_2$Se. There we sample the cross-state matrix elements, which adds a preparation stage before the first query (Appendix~\ref{app:ec-comparison}).

On Li$_2$Se, near-linear dependence among the five anchor states makes the EC eigenproblem sensitive to noise in their sampled overlaps. Each of 16 independent sample batches supplied overlap and Hamiltonian matrices for solving EC at all 16 queries. At the $10^{-6}$ cutoff, independent samples confirmed a positive energy denominator (squared norm) at 95\% confidence for 158 of the 256 states. With 98 unresolved, we averaged the matrices over all 16 batches and solved EC again at each query. All 16 states passed the same check, with an MAE of \result{ec-li2se-1e-06-mae_mha}~mHa against DMRG and energy standard errors up to \result{ec-li2se-1e-06-energy_se_max_mha}~mHa (Table~\ref{tab:ec-li2se}).
Tensor construction for these estimates took \result{ec-li2se-production-setup-hours} eight-thread CPU task-hours, plus 9.08 for the independent validation tensors. Cached EC queries took 0.294~s each including RHF and integral conversion, plus 0.096~s for validation. Sampling the same frozen NNQS directly bypassed the cross-state matrices and took \result{ec-li2se-direct-query-hours} A100 GPU-hours for all 16 queries, plus 0.242~s of CPU preparation per query. Its energy standard errors were at most \result{ec-li2se-direct-energy_se_max_mha}~mHa, and every query reached chemical accuracy against DMRG (MAE \result{ec-li2se-direct-mae_mha}~mHa, maximum error \result{ec-li2se-direct-max_absolute_error_mha}~mHa).

\section{Discussion and conclusion}
We have established geometry-conditioned FNQSs for molecular electronic structure in second quantization. Joint variational training turns sparse anchor calculations into a shared representation of molecular ground states, supplying energies and electronic observables at untrained geometries with frozen parameters. Orbital alignment establishes an aligned orbital basis across geometries, enabling the shared network to learn the physical variation along a molecular path.

The energy-trained states recover charge redistribution and changes in natural occupations without property-specific fitting. At fixed training budgets, denser queries amortize joint training, while per-geometry optimization and evaluation costs determine the asymptotic speedup. This work establishes an efficient route to computing molecular potential energy surfaces with neural-network quantum states in an occupation basis.

\paragraph{Limitations and outlook.}
Our benchmarks cover molecular paths and combined angular and radial distortions of H$_4$, with FCI or DMRG references in specified finite orbital spaces. Basis-set errors remain. Extending the method to larger active spaces will require efficient sampling and Hamiltonian evaluation as the number of relevant configurations grows. In multidimensional geometry domains, orbital matching can depend on the transport path, making consistency around closed loops an additional challenge near orbital crossings. A further direction is to extend orbital correspondence across different active spaces and, ultimately, across molecules. Together with architectures that accommodate varying orbital spaces and electron counts, such correspondences could support pretraining and wavefunction reuse across a broader range of chemical systems.

\subsection*{Reproducibility statement}
Appendices~\ref{app:coordinates}, \ref{app:sampling}, and~\ref{app:protocols} specify orbital alignment and query attachment, network and sampling settings, and molecular paths and electronic sectors, respectively.

\subsection*{AI use statement}
The authors used ChatGPT and Cursor (with Claude Fable 5.1 and Grok 4.7) to assist with method design, experimental design, method implementation and debugging, numerical data processing, and interpretation of results. The tools also assisted with literature search, manuscript editing, and figure preparation. The authors reviewed all AI-assisted work, including the text, code, experimental design, results, and references. The authors take responsibility for the final content of this submission.

\FloatBarrier
\clearpage
\qkbibliography{references}
\clearpage
\nobalance
\qkappendix
\iflatexml
  \section*{Appendices}
\else
  \begin{qkfrontmatter}
  \pdfbookmark[0]{Appendices}{appendices}
  {\centering\fontsize{16}{19}\selectfont\bfseries\color{QKPrimary}Appendices\par}
  \end{qkfrontmatter}
  \thispagestyle{qkbody}
\fi
\numberwithin{equation}{section}
\numberwithin{figure}{section}
\numberwithin{table}{section}
\centerline{\textbf{SUMMARY OF THE APPENDIX}}
\medskip
The appendix is organized as follows:
\begin{itemize}
\item Appendix~\ref{app:coordinates} describes orbital transformations, alignment protocols, and query attachment.
\item Appendix~\ref{app:sampling} specifies the network, optimizer, sampling, and energy-evaluation settings.
\item Appendix~\ref{app:protocols} defines the molecular paths, retained orbital spaces, and electronic observables.
\item Appendix~\ref{app:detailed} reports detailed energy and property results, per-seed ablations, and computational costs.
\item Appendix~\ref{app:additional-energy} examines transfer across a Li$_2$Se bond stretch, combined H$_4$ distortions, and extended N$_2$/CO bond scans.
\end{itemize}

\section{Orbital alignment and query attachment}\label{app:coordinates}
\subsection{Orbital transformations}
At fixed geometry, $\widetilde a_p^\dagger=\sum_q a_q^\dagger U_{qp}$ induces the exterior-power representation of $U$ on determinants. In a basis with separately ordered alpha and beta determinants, its matrix element between occupied sets is
\begin{equation}
\mathcal U_{I,J}=\det U_{I_\alpha,J_\alpha}\det U_{I_\beta,J_\beta}.
\end{equation}
Conversion to interleaved spin-orbital order includes the fermionic ordering signs. Signed permutations therefore relabel determinants with the corresponding parity; general rotations mix them through these minors. For any retained-space unitary, $\mathcal U$ is unitary and
\begin{equation}
(\mathcal U^\dagger\psi)^\dagger(\mathcal U^\dagger H\mathcal U)(\mathcal U^\dagger\psi)=\psi^\dagger H\psi.
\end{equation}
Rotations are confined to the retained orbital space; frozen-core and excluded orbitals define separate boundaries. Across different geometries, the AO functions themselves move, so cross-AO overlaps enter the matching step.

\subsection{Alignment protocols}
The N$_2$/CO construction transports the orbital bases at training anchors from a fixed root, then aligns each query directly to its nearest aligned training anchor. The H$_4$ endpoint RHF solution is continued from the lower canonical branch in the common C$_{2h}$ subgroup.

Orbital preparation also uses query geometries as additional RHF support.

For NH$_3$, orbital preparation resolves parity under a fixed physical reflection, $z\mapsto-z$ in the laboratory frame. The reflection matrix is diagonalized within exactly degenerate Fock blocks (energy tolerance $10^{-7}$~Ha), keeping occupied/virtual and frozen-core boundaries separate. This step rotates degenerate orbitals into a common parity convention before maximum-overlap matching and sign alignment. The orbital bases at the five anchors are transported from the planar root, and each query is attached to its nearest anchor.

\section{Optimization and sampling details}\label{app:sampling}
\subsection{Network and optimizer settings}
Electronic-structure calculations use PySCF \citep{sun2020pyscf}, and neural-network models are implemented in PyTorch \citep{paszke2019pytorch}.

Production models use four amplitude blocks, four heads, and four phase-network layers of width 512 with rectified linear unit activations. The amplitude width is 32 except for ethylene, which uses width 64. Occupations are decoded in reverse orbital order; descriptor and irrep arrays follow that same order. The phase branch concatenates the geometry condition $\mathbf c(\R)$ with the signed spin-occupation vector $\mathbf s(\nn)$. Diatomics use bond lengths in angstroms as geometry features; the other molecular paths rescale their coordinate to $[-1,1]$ using fixed domain bounds. Spin populations are fixed in all runs; the N$_2$ model uses no spatial-irrep mask, while the other five original systems impose the prescribed total irrep. Dropout is zero. The FiLM maps in Equation~\eqref{eq:film} have zero initial weights and biases, with no bias in the descriptor map. The output head applies a final LayerNorm followed by the linear projection. AdamW \citep{loshchilov2019adamw} uses a base learning rate $10^{-3}$, betas $(0.9,0.99)$, epsilon $10^{-9}$, and zero weight decay. Computation uses float32 model parameters.

The six original models use seed 11 and 20,000 updates. N$_2$ uses constant learning rate $10^{-3}$. CO, H$_4$, NH$_3$, and ethylene use cosine decay from $10^{-3}$ to $10^{-5}$ without warmup. CO$_2$ uses $3\times10^{-4}$ through update 6,000, $10^{-4}$ through 15,000, and $3\times10^{-5}$ through 20,000. Training uses two A100 GPUs except for ethylene (four A100s) and NH$_3$, which uses one V100 through update 15,000 and three A100s thereafter. All final models use FiLM after each of the four Transformer blocks. The orbital-alignment and conditioning ablations use the corresponding N$_2$/CO schedules and seeds 11, 23, and 37.

\begin{table*}[t]

\caption{Training budgets and per-geometry distinct-configuration targets.}
\label{tab:workload}
\begin{center}
\begin{tabular}{lrrl}
\toprule
Study & Steps & Geometries/update & Distinct configurations\\
\midrule
N$_2$ & 20,000 & 5 & 6,000--50,000\\
CO & 20,000 & 5 & 600--2,400\\
H$_4$/cc-pVDZ & 20,000 & 4 & 2,000--4,000\\
NH$_3$ & 20,000 & 5 & 6,000--20,000\\
Ethylene & 20,000 & 7 & 6,000--20,000\\
CO$_2$ & 20,000 & 5 & 6,000--20,000\\
Li$_2$Se & 30,000 & 5 & 6,000--50,000\\
\bottomrule
\end{tabular}
\end{center}
\end{table*}

\subsection{Sampled-subspace training and full-state evaluation}
The native Hamiltonian backend constructs in-set single- and double-excitation connections. Network probabilities, normalized over $S_k$, weight its energy and gradient; BAS multiplicities determine set membership. Holding $S_k$ fixed during differentiation gives
\begin{equation}
\begin{aligned}
\nabla_\theta E_{S_k}
={}&2\operatorname{Re}\sum_{x\in S_k}
\frac{|\psi_\theta(x)|^2}{\sum_{y\in S_k}|\psi_\theta(y)|^2}\\
&\quad\times\left[\frac{(H_{S_kS_k}\psi_{S_k})_x}{\psi_\theta(x)}-E_{S_k}\right]
\nabla_\theta\log\psi_\theta(x)^*.
\end{aligned}
\label{eq:gradient}
\end{equation}

The sampled set is refreshed between updates. The initial BAS draw counts per geometry are 32,768 for N$_2$, 2,048 for CO, and 4,096 for H$_4$ and Li$_2$Se. BAS adjusts these counts to the distinct-configuration targets in Table~\ref{tab:workload}. Related sampled-subspace objectives are described by \citet{wu2023sample,malyshev2024}.

In Fig.~\ref{fig:representative-pes} and Appendix~\ref{app:expanded-coverage}, each evaluation replicate adapts its BAS draw count to the distinct-configuration target used during training. Accepted replicates used about $10^{12}$ proposals for both N$_2$ scans, $(7.4\text{--}9.6)\times10^7$ for the five-anchor CO scan and $4.38\times10^7$ for its extension, $(1.57\text{--}2.04)\times10^9$ for H$_4$, and $(0.88\text{--}1.48)\times10^8$ for Li$_2$Se. Repeated configurations are counted together. If $10^{12}$ draws fall short of the target, uniform proposals supplement BAS and $c_x$ includes their importance weights. Otherwise $c_x$ is the sampled multiplicity. The energy estimate is
\begin{equation}
\widehat E=\frac{\sum_x c_x\operatorname{Re}E_{\mathrm{loc}}(x)}{\sum_x c_x}.
\end{equation}
The local-energy sum includes connected states outside $S$, with their amplitudes evaluated by the same frozen network. Three independent runs at each geometry give sampling replicates. The standard error of the mean is the larger of the within-replicate propagated error, $\sqrt{\sum_{r=1}^3\sigma_r^2}/3$, and the between-replicate sample standard deviation divided by $\sqrt{3}$. Relative-energy uncertainties include the point and reference-origin variances in quadrature.

\section{Molecular paths and property conventions}\label{app:protocols}
\begin{table*}[t]

\caption{Molecular systems. Electron/orbital counts refer to the retained space.}
\label{tab:systems}
\begin{center}
\begin{tabular}{llllr}
\toprule
System & Basis & $e^-$/orbitals & Coordinate & Anchors\\
\midrule
N$_2$ & STO-3G & 14/10 & bond length & 5\\
CO & STO-3G & 14/10 & bond length & 5\\
H$_4$ rectangle & cc-pVDZ & 4/20 & rectangle angle & 4\\
NH$_3$ & 6-31G, frozen core & 8/14 & umbrella height & 5\\
CO$_2$ & STO-3G, frozen core & 16/12 & C--O distance & 5\\
C$_2$H$_4$ & STO-3G, frozen core & 12/12 & torsion angle & 7\\
\bottomrule
\end{tabular}
\end{center}
\end{table*}

\paragraph{Diatomics and H$_4$.}
N$_2$ spans 0.60--1.40~\AA{} and CO spans 0.80--1.60~\AA{}, both at 0.02~\AA{} reporting spacing and with five anchors at 0.20~\AA{} spacing. Each grid contains five anchors and 36 non-anchor queries. H$_4$ has coordinates $(\pm x,\pm y,0)$ with $x=R\cos(\theta/2)$, $y=R\sin(\theta/2)$, and $R=3.2843\,a_0$. Its 25-point grid spans $84^\circ$--$90^\circ$ in $0.25^\circ$ steps, with anchors $84^\circ,86^\circ,88^\circ,90^\circ$, and 21 non-anchor queries. The C$_{2h}$/$A_g$ convention continues the RHF density from $89.75^\circ$ to the square endpoint.

\paragraph{NH$_3$.}
The coordinate $q$ is the nitrogen height above the H$_3$ plane at fixed N--H distance 1.02~\AA. The five anchors are $q=0,0.15,0.30,0.45,0.60$~\AA, within a uniform 25-point reporting grid. A common vertical C$_s$ reflection is used at every geometry, including the planar endpoint. N~$1s$ is frozen, leaving eight electrons in 14 orbitals. The dipole axis points from the H$_3$ plane toward positive N height.

\paragraph{CO$_2$.}
The molecule is linear with carbon at the origin and oxygen atoms at $z=\pm r$, with $r\in[1.05,1.35]$~\AA. Five anchors at 0.075~\AA{} spacing define the aligned orbital bases; the 25-point reporting grid has 0.0125~\AA{} spacing. Matching preserves D$_{2h}$ irreps and the core, occupied, and virtual orbital classes. Freezing the C and O $1s$ cores leaves 16 electrons in 12 STO-3G orbitals. The reported traceless axial quadrupole is $Q_{zz}=\tfrac12\sum_a q_a(3z_a^2-r_a^2)$, including electronic, frozen-core, and nuclear contributions, in $e a_0^2$.

\paragraph{Ethylene.}
The fixed planar structure has C=C distance 1.339~\AA, C--H distance 1.087~\AA, and C--C--H angle $121.3^\circ$. Symmetric fragment rotations by $\pm\theta/2$ define $\theta\in[0,90^\circ]$. Seven Chebyshev--Lobatto anchors are $0,6.028857,22.5,45,67.5,83.971143,90^\circ$. The reporting grid comprises 33 uniformly spaced angles and the two additional anchors at $6.028857^\circ$ and $83.971143^\circ$, giving 35 points and 28 queries. All seven anchors share the orbital labels and phase conventions of the initial five-anchor basis. Both carbon $1s$ cores are frozen, leaving 12 electrons in 12 orbitals in a common C$_2$ subgroup. The displayed frontier occupations are the sixth and seventh eigenvalues of the active spin-summed 1-RDM, sorted in descending order; their $\pi/\pi^*$ character is checked against the reference natural orbitals.

\paragraph{Electronic observables.}
The spin-summed 1-RDM is Hermitized before diagonalization, and its trace equals the retained electron count. Natural occupations are sorted in descending order. Dipoles include electronic, frozen-core, and nuclear contributions with a fixed origin convention. Property MAEs use all non-anchor reporting points; ethylene metrics average occupations 6 and 7. RHF and CCSD \citep{purvis1982ccsd} use the same geometries, retained spaces, origins, and operators as NQS and FCI. CCSD properties are obtained by contracting the relevant operators with the unrelaxed spin-summed one-particle density matrix from converged amplitude and $\Lambda$ equations. Occupied and virtual orbitals are separately canonicalized while retaining the RHF determinant. All reporting geometries have converged solutions.

\FloatBarrier
\section{Supporting results}\label{app:detailed}
\subsection{Frozen energy accuracy}\label{sec:frozen}
\begin{figure*}[t]
\centering\includegraphics[width=\linewidth]{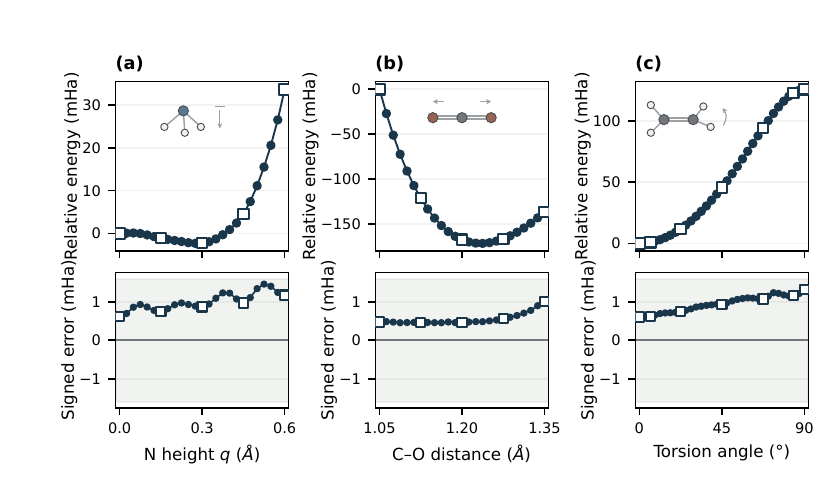}
\caption{\textbf{Frozen energy predictions.} (a) NH$_3$ inversion, (b) CO$_2$ symmetric stretching, and (c) ethylene torsion. Each panel combines frozen NQS energies relative to each path origin (upper) and the corresponding signed complete-Hamiltonian errors against FCI (lower), with the $\pm1.6$~mHa band. Hollow squares mark anchors; dots mark queries. Error bars show one sampling standard error. The corresponding property panels appear in Fig.~\ref{fig:molecular-properties}.}
\label{fig:paths}
\end{figure*}

Table~\ref{tab:overview} reports the anchor and query errors for Fig.~\ref{fig:representative-pes}. Table~\ref{tab:all-energies} and Fig.~\ref{fig:paths} give the energies of the same checkpoints used for the three property panels. Every query in Table~\ref{tab:overview} is within 1.6~mHa of its reference energy.

\begin{table*}[t]

\caption{Frozen molecular predictions, in mHa relative to FCI (DMRG for Li$_2$Se). Anchor and query metrics use the same final checkpoint.}
\label{tab:overview}
\begin{center}
\begin{tabular}{llrrrr}
\toprule
System & Molecular path & Anchors / queries & Anchor MAE & Query MAE & Query max.\\
\midrule
N$_2$ & bond stretch & 5 / 36 & \result{prod-n2-mae} & \result{final-n2-query-mae} & \result{final-n2-query-max}\\
CO & bond stretch & 5 / 36 & \result{prod-co-mae} & \result{final-co-query-mae} & \result{final-co-query-max}\\
H$_4$ & rectangle & 4 / 21 & \result{prod-h4-mae} & \result{final-h4-query-mae} & \result{final-h4-query-max}\\
Li$_2$Se & bond stretch & 5 / 16 & \result{fig2-li2se-anchors-mae_mha} & \result{fig2-li2se-queries-mae_mha} & \result{fig2-li2se-queries-maximum_absolute_error_mha}\\
\bottomrule
\end{tabular}
\end{center}
\end{table*}

\subsection{Molecular-property benchmarks}\label{sec:paths}
\begin{table*}[t]

\caption{Energy errors for the three molecular-property benchmarks in Fig.~\ref{fig:molecular-properties}, in mHa relative to FCI. Anchor and query metrics use the same final checkpoints as the property comparisons; queries include all non-anchor grid points.}
\label{tab:all-energies}
\begin{center}
\begin{tabular}{llrrrr}
\toprule
System & Molecular path & Anchors / queries & Anchor MAE & Query MAE & Query max.\\
\midrule
C$_2$H$_4$ & torsion & 7 / 28 & \result{prod-c2h4-mae} & \result{final-c2h4-query-mae} & \result{final-c2h4-query-max}\\
NH$_3$ & inversion & 5 / 20 & \result{prod-nh3-mae} & \result{final-nh3-query-mae} & \result{final-nh3-query-max}\\
CO$_2$ & symmetric stretch & 5 / 20 & \result{prod-co2-mae} & \result{final-co2-query-mae} & \result{final-co2-query-max}\\
\bottomrule
\end{tabular}
\end{center}
\end{table*}

Table~\ref{tab:property-baselines} quantifies the property comparisons in Fig.~\ref{fig:molecular-properties}. Each metric uses all non-anchor reporting points; occupation errors average over both displayed eigenvalues.

\begin{table}[tb]
\caption{Query property MAEs against FCI. All methods use identical geometries, retained spaces, and observable conventions.}
\label{tab:property-baselines}
\begin{center}
\begin{tabular}{lrrr}
\toprule
Observable & RHF & CCSD & NQS\\
\midrule
NH$_3$ dipole (D) & 0.06057 & 0.00575 & \result{nh3-dipole-mae}\\
CO$_2$ quadrupole ($e a_0^2$) & 1.07714 & 0.06971 & \result{co2-quadrupole-mae}\\
C$_2$H$_4$ occupations & 0.27589 & 0.00485 & \result{c2h4-occupation-mae}\\
\bottomrule
\end{tabular}
\end{center}
\end{table}

\subsection{Per-seed component ablations}\label{sec:mechanism}
\begin{figure*}[t]
\centering\includegraphics[width=\linewidth]{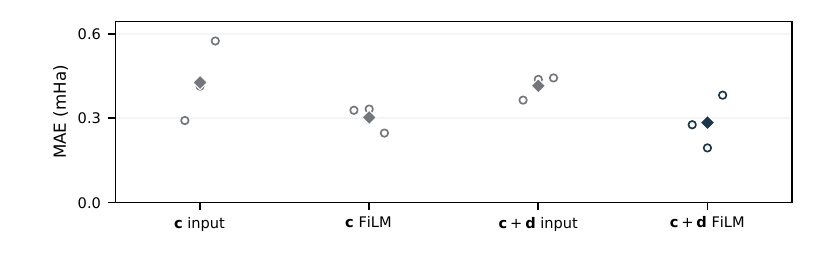}
\caption{\textbf{Conditioning ablations.} All variants use the aligned CO orbital basis. Open circles show query MAEs for three seeds and filled diamonds their means. The geometry condition $\mathbf c$ and orbital descriptors $\mathbf d$ enter either the input embeddings or post-block FiLM adapters. The backbone, phase network, training budget, and evaluation geometries are matched.}
\label{fig:conditioning-ablation}
\end{figure*}

Table~\ref{tab:current-ablations} gives the individual runs underlying Figs.~\ref{fig:component} and~\ref{fig:conditioning-ablation}. Alignment improves N$_2$ query accuracy for every paired seed. All four CO conditioning variants reach chemical accuracy; their differences are small compared with the orbital-alignment effect. The Transformer and phase-network backbones, sampler settings, and optimization budgets are matched within each comparison, with seeds 11, 23, and 37.

\paragraph{Conditioning architecture in the aligned orbital basis.}
In the aligned orbital basis, Fig.~\ref{fig:conditioning-ablation} compares four ways to condition the amplitude. The $\mathbf c$ input baseline adds an affine embedding of the geometry condition $\mathbf c$ (the bond length for CO) to the token and position embeddings; $\mathbf c+\mathbf d$ input additionally supplies a linear projection of the orbital descriptor $d_j$. Their FiLM counterparts apply these signals after each Transformer block, as in Equation~\eqref{eq:film}. Final 20,000-update checkpoints are evaluated on eight CO query geometries over three seeds.

FiLM lowers the mean query MAE from 0.427 to 0.303~mHa with geometry alone. Descriptors have a smaller effect at the input, but combining them with FiLM gives the lowest mean MAE, 0.284~mHa. These differences are comparable to seed variation and smaller than the alignment effect. We use $\mathbf c+\mathbf d$ FiLM in the final model.

\begin{table}[!htbp]

\caption{Per-seed component ablations (MAE in mHa). N$_2$ uses all 36 query geometries; CO uses the eight query geometries. All runs use 20,000 updates with post-block FiLM where applicable; N$_2$ uses a fixed learning rate of $10^{-3}$ and CO uses cosine decay. Each cell reports the error of a fixed final checkpoint.}
\label{tab:current-ablations}
\begin{center}
\begin{tabular}{lrrr}
\toprule
Model & Seed 11 & Seed 23 & Seed 37\\
\midrule
N$_2$ unaligned & \result{current-n2-raw-seed11} & \result{current-n2-raw-seed23} & \result{current-n2-raw-seed37}\\
N$_2$ aligned & \result{current-n2-aligned-seed11} & \result{current-n2-aligned-seed23} & \result{current-n2-aligned-seed37}\\
CO $\mathbf c$ input & \result{current-co-r-input-seed11} & \result{current-co-r-input-seed23} & \result{current-co-r-input-seed37}\\
CO $\mathbf c$ FiLM & \result{current-co-r-film-seed11} & \result{current-co-r-film-seed23} & \result{current-co-r-film-seed37}\\
CO $\mathbf c+\mathbf d$ input & \result{current-co-r-energy-input-seed11} & \result{current-co-r-energy-input-seed23} & \result{current-co-r-energy-input-seed37}\\
CO $\mathbf c+\mathbf d$ FiLM & \result{current-co-r-energy-film-seed11} & \result{current-co-r-energy-film-seed23} & \result{current-co-r-energy-film-seed37}\\
\bottomrule
\end{tabular}
\end{center}
\end{table}

\subsection{Computational cost protocol and estimates}\label{sec:baselines}
We compare four fixed-budget workflows on the N$_2$ bond scan. Table~\ref{tab:cost} gives the budgets, MAEs on the same eight geometries, and total cost estimates. All energies use three independent full-Hamiltonian evaluation replicates. The cost checkpoints are separate from the 20,000-update models used in the main energy results.

\begin{table*}[t]
\setlength{\tabcolsep}{4pt}
\caption{Fixed-budget N$_2$ comparison. MAEs use the common eight geometries. Costs include training and final energy evaluation; independent and shared costs are estimated from the sampled geometries. The two warm-start MAEs correspond to the 41- and 161-point grids, respectively.}
\label{tab:cost}
\begin{center}
\begin{tabular}{llrrr}
\toprule
Workflow & Updates & MAE (mHa) & \shortstack{41 points\\(GPU-h)} & \shortstack{161 points\\(GPU-h)}\\
\midrule
Independent & 4,000 & 1.027 & 10.785 & 42.351\\
Shared + fine-tune & 2,000 + 200/point & 1.064 & 1.682 & 4.162\\
Sequential warm-start & 3,000 + 100/next point & 1.173 / 0.988 & 0.664 & 1.948\\
Joint + frozen & 4,000 & 0.793 & \result{cost-gpu-frozen-41} & 1.644\\
\bottomrule
\end{tabular}
\end{center}
\end{table*}

\paragraph{Training budgets.}
Independent optimization starts each geometry from random initialization and uses AdamW for 4,000 updates, with 100 updates of linear warmup to $3\times10^{-4}$ and cosine decay to $3\times10^{-5}$ at update 4,000. Joint training uses five anchors and a constant learning rate of $10^{-3}$ for 4,000 updates. Shared pretraining removes geometry conditioning and minimizes the equally weighted mean of the five anchor losses for 2,000 updates at $3\times10^{-4}$. Each query then receives 200 fine-tuning updates with a fresh optimizer. Sequential warm starts optimize the first geometry for 3,000 updates, then transfer model weights to the next geometry with a fresh optimizer for 100 updates. The first-point schedule uses cosine decay from $10^{-3}$ toward $10^{-5}$ over 20,000 updates, without warmup. Shared fine-tuning and subsequent warm-start points use 200-update linear warmup toward $10^{-4}$ and cosine decay toward $10^{-5}$ over 10,000 updates. All workflows use the same aligned orbital basis and full-Hamiltonian evaluation protocol.

\paragraph{Hardware and accounting.}
Independent optimization and adaptation use one A100; joint training and shared pretraining use two. GPU-hours sum training, process initialization, checkpoint loading, and final full-Hamiltonian evaluation across allocated GPUs; CPU orbital and integral preparation is outside this GPU-cost accounting. Shared pretraining and joint training cost 0.835 and 1.601 GPU-hours, respectively. The frozen-evaluation slope in Fig.~\ref{fig:cost-scaling} is 0.961 GPU-seconds per geometry, measured over the 161-point grid with three sampling replicates per geometry. Frozen evaluation uses one A100.

RHF, orbital alignment, and integral preparation on eight CPU threads take 1.87~s for 41 points and 6.91~s for 161. Including this stage once, joint training with frozen queries and sequential warm starts take \result{cost-full-time-frozen-41} and \result{cost-full-time-warm-41}~h on the sparse grid, and \result{cost-full-time-frozen-161} and \result{cost-full-time-warm-161}~h on the dense grid, respectively. CPU preparation accounts for 0.23\% and 0.10\% of the dense-grid elapsed times. Training and energy evaluation dominate both.

\paragraph{Timing geometries and accuracy.}
Independent optimization and shared fine-tuning are timed at the same eight geometries selected by stratified sampling (Table~\ref{tab:cost-points}); per-geometry costs are weighted by the fraction of the grid represented by each stratum. All four workflows report MAE on these common points. Sequential warm starts traverse the 41- and 161-point grids in ascending bond length, with full-grid MAEs of 1.182 and 0.993~mHa; frozen evaluation gives 0.823 and 0.822~mHa, respectively. At these attained errors, warm starts cost fewer GPU-hours on the sparse grid and joint training saves 16\% on the dense grid.

\begin{table*}[t]

\caption{Absolute full-Hamiltonian energy errors (mHa) at the eight common N$_2$ comparison geometries. Warm-start columns correspond to the two complete sequential paths.}
\label{tab:cost-points}\label{tab:shared-cost-points}
\begin{center}
\begin{tabular}{rrrrrr}
\toprule
$R$ (\AA) & Independent & Shared + fine-tune & Warm (41) & Warm (161) & Joint + frozen\\
\midrule
0.60 & 0.284 & 0.458 & 0.365 & 0.365 & 0.327\\
0.76 & 0.663 & 0.473 & 0.559 & 0.458 & 0.527\\
0.84 & 1.054 & 0.601 & 0.705 & 0.566 & 0.573\\
0.96 & 1.097 & 0.728 & 0.898 & 0.787 & 0.723\\
1.00 & 0.709 & 0.790 & 0.998 & 0.872 & 0.788\\
1.18 & 1.868 & 1.477 & 1.666 & 1.392 & 1.088\\
1.28 & 1.755 & 1.864 & 2.019 & 1.669 & 1.138\\
1.32 & 0.782 & 2.123 & 2.176 & 1.795 & 1.176\\
\bottomrule
\end{tabular}
\end{center}
\end{table*}

\paragraph{Query-count scaling.}
For each workflow, Equation~\eqref{eq:surface-speedup} holds its measured average per-geometry cost fixed. For warm starts, the intercept is the first-point training and evaluation cost minus one subsequent-point cost, so that $G=1$ recovers the first-point cost exactly. The two warm-start curves use the mean costs measured at their respective grid spacings. Writing the joint-training cost as $a_{\mathrm{frozen}}$ and the per-geometry evaluation cost as $b_{\mathrm{frozen}}$, the estimated speedup over joint training with frozen evaluation is
\begin{equation}
S_m(G)=\frac{a_m+G b_m}{a_{\mathrm{frozen}}+G b_{\mathrm{frozen}}},\qquad
\lim_{G\to\infty}S_m(G)=\frac{b_m}{b_{\mathrm{frozen}}}.
\label{eq:amortization-limit}
\end{equation}
The slope ratios are 986 for independent optimization, 77.4 for shared fine-tuning, and 40.8--43.0 for sequential warm starts. The 41- and 161-point grids have been evaluated for joint and sequential workflows; curves at other query counts are cost-model projections. Equating the two total costs at the measured warm-start rates gives crossings at 126 and 132 geometries for the 41- and 161-point spacings. Thus about 130 geometries, including anchors, cover the initial cost of joint training on this scan.

\subsection{Eigenvector continuation}\label{app:ec-comparison}
Direct evaluation uses the frozen NNQS at the query geometry. For eigenvector continuation (EC), we evaluate the same network at the five training geometries and project the query Hamiltonian into the span of those wavefunctions. For N$_2$ and CO, we can enumerate all 14,400 coefficients per anchor and transform them from the aligned HF orbitals to consistently indexed L\"owdin-orthogonalized atomic orbitals \citep{rath2025}. Overlaps then reduce to inner products. Enumerating the five anchors takes 3.36~s for N$_2$ and 5.88~s for CO. Once their transition densities are stored, contraction and the eigensolve take 1.29 and 1.61~ms per query. RHF and integral conversion account for most of the query time in Table~\ref{tab:ec-li2se}.

\begin{table*}[t]
\centering
\setlength{\tabcolsep}{6pt}
\caption{Costs and attained errors for direct evaluation and EC on the same query geometries. EC matrix preparation counts transition-density kernels on eight CPU threads, summed over anchor pairs for Li$_2$Se. EC query times include RHF and integral conversion. The direct and EC errors use full vectors for N$_2$ and CO (36 queries each) and sampling for Li$_2$Se (16 queries).}
\label{tab:ec-li2se}
\begin{tabular}{lrrrrr}
\toprule
System & \shortstack{Shared training\\(GPU-h)} & \shortstack{EC matrix\\preparation} & \shortstack{EC query\\(s)} & \shortstack{Direct MAE\\(mHa)} & \shortstack{EC MAE\\(mHa)}\\
\midrule
N$_2$     & 7.66   & 2.42 s & 0.116 & 0.0495 & 0.0329\\
CO        & 6.10   & 3.04 s & 0.138 & 0.2528 & 0.1535\\
Li$_2$Se  & 63.474 & \result{ec-li2se-production-setup-hours} task-h & 0.294 & \result{ec-li2se-direct-mae_mha} & \result{ec-li2se-1e-06-mae_mha}\\
\bottomrule
\end{tabular}
\end{table*}

With $9.66\times10^{12}$ determinants per Li$_2$Se anchor, enumeration is impractical. We sample each anchor in its own aligned HF orbitals and use nonorthogonal determinant matrix elements in common atomic-orbital coordinates to estimate the cross-state tensors. Nearly dependent anchors make the overlap matrix sensitive to sampling noise. At a relative overlap cutoff of $10^{-6}$, only 158 of 256 individual fit--query pairs had a positive lower 95\% confidence bound on the norm. Averaging the matrices from 16 fits gave positive bounds at all 16 queries. Independent validation adds 9.08 eight-thread CPU task-hours for tensor construction and 0.096~s per query.

Once these tensors are stored, contraction and the eigensolve take \result{ec-li2se-query-seconds}~s of the 0.294~s per Li$_2$Se query. Direct evaluation samples anew at each geometry. Using the checkpoint of Fig.~\ref{fig:representative-pes}(d), the timed run took \result{ec-li2se-direct-query-hours} A100 GPU-hours across the 16 queries (six replicates of 32,768 draws each), plus 0.242~s of CPU preparation per geometry. Further EC queries along the same scan reuse the tensors, spreading their construction cost over more geometries.

\FloatBarrier
\section{System size and geometric transfer}\label{app:additional-energy}
\subsection{\texorpdfstring{Li$_2$Se}{Li2Se} beyond enumerated FCI spaces}\label{app:li2se}
Stretching both Li--Se bonds of linear Li$_2$Se together tests whether the accuracy at the anchors carries over to new geometries in a much larger determinant space. The all-electron STO-3G calculation has 40 electrons in 28 spatial orbitals, or $\binom{28}{20}^2\simeq9.66\times10^{12}$ determinants. We trained at five bond lengths spaced by 0.10~\AA{} from 2.20 to 2.60~\AA. We test at the remaining 16 positions on a 0.02~\AA{} grid.

We computed DMRG reference energies with block2 \citep{white1992dmrg,zhai2023block2}. Increasing the bond dimension from 1,024 to 1,536 changed the energy by at most $2.30\times10^{-7}$~Ha across the scan. The mean absolute error is \result{fig2-li2se-anchors-mae_mha}~mHa at the anchors and \result{fig2-li2se-queries-mae_mha}~mHa at the 16 queries (Table~\ref{tab:overview}). Interpolation therefore adds little to the mean error on this stretch. The largest query error is \result{fig2-li2se-queries-maximum_absolute_error_mha}~mHa.

\subsection{\texorpdfstring{H$_4$}{H4} transfer to combined geometric distortions}\label{app:h4-composition}
We first evaluated the original angle-only H$_4$ model at six geometries off its training path. At $\theta=87^\circ$, changing the radius by $\pm1\%$, $\pm3\%$, and $\pm5\%$ with the weights frozen gives an MAE of 0.373~mHa and a maximum error of 0.829~mHa. Matched orbital overlaps remain at least 0.966. At this angle the model retains chemical accuracy as the radius varies along a direction held fixed throughout training.

Figure~\ref{fig:geometric-transfer} tests whether a model can combine distortions it encountered separately. We express angle and radius as $x=(\theta-87^\circ)/2^\circ$ and $y=(r/r_0-1)/0.05$, with $r_0=3.2843\,a_0$ (Appendix~\ref{app:protocols}). One model trains at the center and at $\pm0.5$ and $\pm1$ along each axis. A control uses nine anchors on the tensor grid $\{-1,0,1\}^2$. Both receive 20,000 updates and are evaluated on the same 36 combinations $\{\pm0.25,\pm0.50,\pm0.75\}^2$.

With axis-only training, MAE rises from \result{additional-h4-axis-interior-mae_mha}~mHa inside the training hull to \result{additional-h4-axis-exterior-mae_mha}~mHa outside it (Table~\ref{tab:h4-composition}). Across all 36 queries, axis-only and tensor-grid training give MAEs of \result{additional-h4-axis-mae_mha} and \result{additional-h4-grid-mae_mha}~mHa, respectively. The 12 points outside the diamond in Fig.~\ref{fig:geometric-transfer}(b) still lie within the training range of each individual coordinate. Their errors therefore measure how well the model combines familiar angular and radial displacements.

\begin{table*}[t]
\centering
\setlength{\tabcolsep}{5pt}
\caption{H$_4$ combined-distortion query errors against FCI, in mHa. Interior, boundary, and exterior refer to the axis-only training hull $|x|+|y|\leq1$.}
\label{tab:h4-composition}
\begin{tabular}{lrrrrr}
\toprule
 & & \multicolumn{2}{c}{Axis-only training} & \multicolumn{2}{c}{Tensor-grid training}\\
Queries & Count & MAE & Maximum & MAE & Maximum\\
\midrule
Interior & 12 & \result{additional-h4-axis-interior-mae_mha} & \result{additional-h4-axis-interior-maximum_absolute_error_mha} & \result{additional-h4-grid-interior-mae_mha} & \result{additional-h4-grid-interior-maximum_absolute_error_mha}\\
Boundary & 12 & \result{additional-h4-axis-boundary-mae_mha} & \result{additional-h4-axis-boundary-maximum_absolute_error_mha} & \result{additional-h4-grid-boundary-mae_mha} & \result{additional-h4-grid-boundary-maximum_absolute_error_mha}\\
Exterior & 12 & \result{additional-h4-axis-exterior-mae_mha} & \result{additional-h4-axis-exterior-maximum_absolute_error_mha} & \result{additional-h4-grid-exterior-mae_mha} & \result{additional-h4-grid-exterior-maximum_absolute_error_mha}\\
All & 36 & \result{additional-h4-axis-mae_mha} & \result{additional-h4-axis-maximum_absolute_error_mha} & \result{additional-h4-grid-mae_mha} & \result{additional-h4-grid-maximum_absolute_error_mha}\\
\bottomrule
\end{tabular}
\end{table*}

\subsection{Extended bond scans}\label{app:expanded-coverage}
\begin{figure*}[t]
\centering\includegraphics[width=\linewidth]{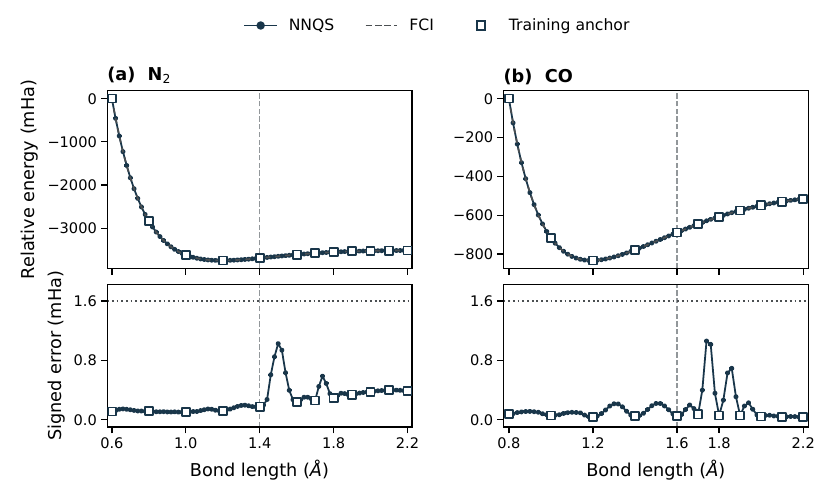}
\caption{\textbf{Extended bond scans.} (a) N$_2$ and (b) CO through 2.20~\AA. Upper panels show frozen NNQS and FCI energies relative to their respective first geometries. Lower panels show signed NNQS errors against FCI, with the dotted line at 1.6~mHa. Open squares mark the 12 and 11 training points, dots mark test points, and error bars show one sampling standard error. Vertical dashed lines mark the upper bounds of the original five-anchor scans.}
\label{fig:extended-bonds}
\end{figure*}

Figure~\ref{fig:extended-bonds} shows separate models trained over intervals extending to 2.20~\AA. We refined the 0.20~\AA{} training grid from 1.60 to 2.20~\AA{}, where the RHF HOMO--LUMO gaps are smaller than at shorter bond lengths. Adding points at 1.70, 1.90, and 2.10~\AA{} gives 12 training points for N$_2$ and 11 for CO. We test the frozen models at the remaining 69 and 60 points, respectively, on the 0.02~\AA{} grids. The test MAEs are \result{fig2-n2-queries-mae_mha} and \result{fig2-co-queries-mae_mha}~mHa, respectively. The largest test errors are \result{fig2-n2-queries-maximum_absolute_error_mha}~mHa at 1.50~\AA{} for N$_2$ and \result{fig2-co-queries-maximum_absolute_error_mha}~mHa at 1.74~\AA{} for CO. Both are below 1.6~mHa.

\end{document}